\documentclass[a4paper,twoside,11pt]{article}
\usepackage[pdftex]{graphicx}
\usepackage{authblk}
\usepackage{amsmath,amsfonts,amsthm,amssymb,bm}
\usepackage{subfigure,enumerate}
\title{TCP over 3G links: Problems and Solutions}
\author{Ioannis Koukoutsidis\\ \texttt{email: i.koukoutsidis@di.uoa.gr}}
\date{}
\begin{document}
\maketitle
\begin{centering}
\begin{abstract}
This review paper presents analytical information regarding the transfer
of TCP data flows on paths towards interconnected wireless systems,
with emphasis on 3G cellular networks. The focus is on protocol
modifications in face of problems arising from terminal mobility
and wireless transmission.

The objective of this paper is not to present an exhaustive
review of the literature, but to filter out the causes of poor TCP
performance in such systems and give a rationalized view of
measures that can be taken against them.
\end{abstract}
\end{centering}
%
%
\section{Introduction}
The proliferation of the Transmission Control Protocol (TCP) in
Internet communications today incites the research community to
further extend its use in mobile and wireless networks. The
ultimate goal is efficient and reliable TCP flows for Internet
traffic over interconnected wired and wireless paths, where the
wireless path suffers from additional problems due to higher BERs
(Bit Error Rates) and frequent link changes. This primarily
entails the treatment of protocol issues, but also additional
inter-operability in the network infrastructure.

In the large-scale mobility case, cellular networks of the
$3^{\textrm{rd}}$ generation (3G) are the most suitable candidates
for support of Internet traffic, as they offer capacity for
enhanced broadband data transfers, as well as improved
transmission quality. They are predominantly characterized by CDMA
(Code Division Multiple Access) transmission technology
\cite{Korhonen03}.

\begin{figure}[!htp]
    \centering
    \includegraphics[scale=1]{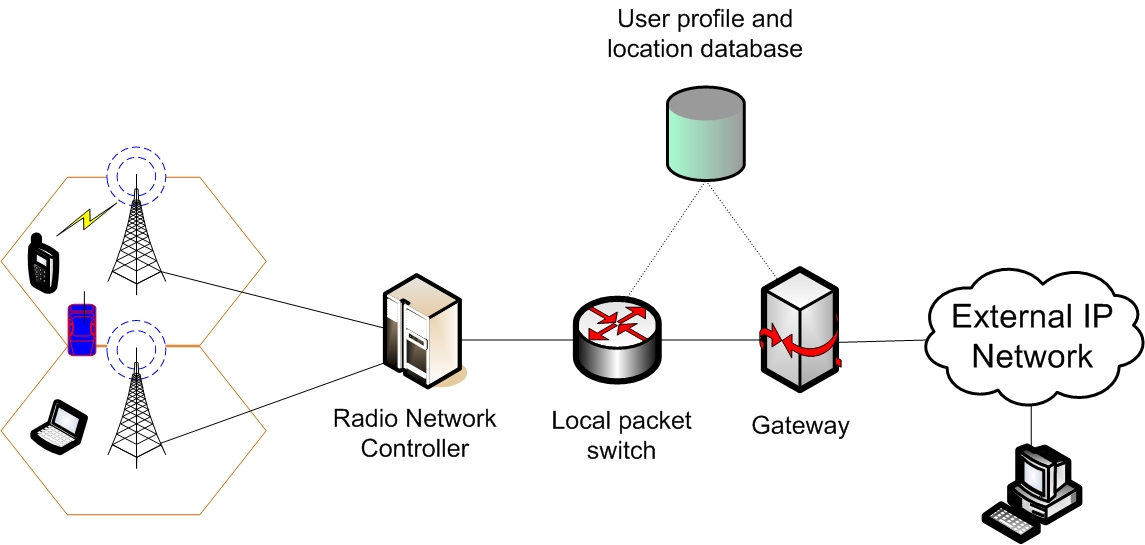}
    \caption{3G network architecture for IP communications}
    \label{fig1}
\end{figure}

Figure~\ref{fig1} portrays the typical 3G network architecture for
IP communications. The radio network controller (RNC) manages several base station (BS)
transceivers and is responsible for handover operations. Packets
are routed through a local switch, while a gateway switch ensures
the connection to an external IP network. User profile and
location information are maintained at a separate database.

Although the focus in this review is on cellular networks, many of
the issues treated apply generally to wireless access systems.
\section{TCP problems in 3G links}
The transfer of Internet packet flows in wireless paths requires
TCP to also be capable of handling transmission losses and
discerning them from losses due to congestion. TCP was designed
for wireline networks, where transmission losses are extremely low
(BERs in the order of $10^{-10}$, and down to $10^{-12}$ for
optical links). Wireless links, on the other hand, may experience
much higher BERs due to interference and fading phenomena
(although care is taken for loss rates not to exceed $10^{-2}$).
Such random losses, if perceived as congestion, will lead to
spurious timeouts or mistakenly trigger the TCP sender to reduce
its congestion window and hence its sending rate unnecessarily.
Problems then manifest themselves as degradation of throughput, as
well as inefficiency in network link utilization. Moreover, the
fact that random bit errors usually occur in short bursts leads to
a higher probability of multiple packet (or segment, in initial
TCP terminology) losses within one transmission window, further
degenerating this behavior: bursty packet losses additionally make
the system get stuck in lengthy recovery procedures (and most
probably hopeless, since timeouts may occur anyhow), especially
when selective or partial acknowledgements are not supported. 

Problems for TCP in cellular networks also occur during the
handover procedure, when a mobile moves to the coverage area of
different base stations (BS). Generally, and as it will be elaborated in a later section,
handovers induce temporal disconnections, buffer losses, increased latency and packet reordering.
In CDMA, universal frequency reuse makes it possible to easily
perform soft handoff, whereupon a mobile is connected to two or more
BS. This can eliminate temporal signal loss or the need to suspend
and resume the transmission of packets, which --if no notification
is sent to the TCP sender-- may result in buffer overflow.
However, problems may still occur if soft handoff is not initiated
promptly or if power control is not properly configured.

Additionally, TCP transmissions in 3G CDMA links are generally
struck by increased latency. This is due to the extensive
processing required at the physical layer of these links for
coding and interleaving, and to link layer processing for FEC
(Forward Error Control) and link-level retransmissions. Further,
since a 3G link is frequently assigned to a single host, there is
a low degree of statistical multiplexing and RTT variations occur
more easily.

Spurious retransmissions and timeouts may also occur because of
dynamic resource sharing schemes in 3G links. Variable rate
support in CDMA has the potential to offer, apart from plain
multiservice traffic, dynamic resource scheduling and link
utilization (e.g., the HSDPA scheme in WCDMA and HDR in cdma2000
\cite{Parkvall01,Bender00}). However, the temporal allocation and
deallocation of resources can cause delay jitter, blocking and
timeouts.

Furthermore, it is very difficult to achieve good overall TCP performance with variable rate support.
Users assigned very high data rates may not reach the bandwidth-delay product, while
low rate users are faced with overflow problems (there is limited memory available for the TCP/IP stack).

Link asymmetry is also an issue of concern for TCP design. Specifically, transmission
rates in the downlink are usually much higher than in the uplink, particularly for
the high speed HSDPA and HDR implementations. Even if this asymmetry can be tolerated
by TCP's self clocking mechanism, there is a much higher risk of buffer overflows both at the 
receiving and transmitting side. 
Ragged behavior will also evidently occur if poor transmission conditions weigh more 
in one direction than the other.

Finally, another handicap is that TCP/IP header compression techniques,
which reduce data traffic load, do not perform properly on
wireless links. Specifically, most of these schemes do not
transmit the entire TCP/IP header, but only changes in consecutive
headers (e.g., the RFC 1144 header compression algorithm \cite{Jacobson90}). Upon
frequent TCP losses, the transmitter and receiver may fall out of
synchronization, and continuously discard packets because of
checksum errors.

\vspace{6pt} {\em For a more complete understanding of problematic
TCP behavior, basic TCP principles should be revised
(\cite{Stevens94} or at least
\cite{Jacobson88,Stevens97,Postel81}). Readers can also consult
the related RFC \cite{Inamura03} or more general review papers
such as \cite{Tian05,Thoppian}.}

\vspace{6pt}

The improvement of TCP performance is a matter of the overall system design,
protocol issues, architecture and communication of network entities, buffer management, 
as well as packet scheduling algorithms. Here we concentrate on 
the modification of the data communication protocol suite, which is the primary issue and
should aim at the treatment or prevention of degenerative effects.
Solutions can be loosely classified into link layer mechanisms,
end-to-end TCP modifications and split-end approaches. Below is a
critical view at solution approaches, rather than solution
schemes. For more detailed scheme descriptions readers can refer
to \cite{Tian05,Thoppian} and the references therein.
\section{Link layer mechanisms}
\begin{figure}[!htp]
    \centering
    \includegraphics[scale=1]{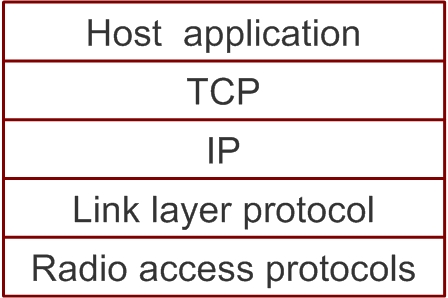}
    \caption{Data communication protocol stack with addition of link layer in 3G systems.
    Radio access protocols provide resource control and medium access
    (e.g., in UMTS these are composed of RRC, MAC and PHY layer
    protocols).}
    \label{fig2}
\end{figure}

A link layer is used in communication protocols to provide flow
control, as well as error detection and correction. Essentially,
this is the task performed by TCP.\footnote{The Internet layer
architecture is older than the OSI reference system. The final
recommendation for TCP and IP was done in 1981, while the OSI
reference model came out in 1983.} Nevertheless, the addition of a
link layer operating underneath the TCP/IP stack (see Figure
\ref{fig2}) can be very useful in wireless paths, and is adopted in 3G access systems.

Link layer retransmission mechanisms enable more prompt recovery
from wireless losses. Their primary mission should be to reduce
the packet error rate to a level that would not cause significant
performance degradation at the TCP layer, which will in turn
eliminate all detectable errors. In general, it would be desired
to limit the packet error rate to the same level of a wireline
network (below $10^{-8}$).

A link layer can also assume the role of fragmenting (and
reassembling) packets into smaller segments, more suitable for
transmission over a wireless link (since a smaller-sized packet
leads to smaller error probability).

Currently a negative acknowledgement ARQ mechanism is specified in
cdma2000 \cite{3GPP2}, while a more complex mechanism with status
reports containing both received and missing packets is specified
in UMTS/WCDMA \cite{3GPP}. Different ARQ mechanisms may
further be implemented in high data rate schemes (e.g., hybrid ARQ in HSDPA \cite{Parkvall01}). 

However, one must keep in mind that the link layer solution does
not completely shield the TCP sender. TCP performance can still be
poor for two reasons:
\begin{itemize}
\item retransmit timeouts caused by additional link layer delay
\item out of order reception of data from the link layer, leading
to unnecessary invocations of the TCP fast retransmit mechanism
\end{itemize}
Both delay and out-of-order delivery can result in competing
(duplicate) retransmissions from the two layers, and thus
inefficient utilization of the network capacity. To cope with
this, a link layer implementation should be ``TCP-aware'', meaning
to have the capability to view TCP headers and understand TCP
semantics. Then it could be equipped with more ``intelligent''
functions, such as holding back duplicate acknowledgements that
would trigger fast retransmissions, or control the setting of the
retransmit timer at the TCP layer.
\section{End-to-end TCP modifications}
TCP is a connection-oriented end-to-end protocol and therefore
solution approaches should primarily aim at protocol modifications
at the end hosts.
General protocol improvements from experimental methods are listed below.
\begin{itemize}
\item[-] The use of selective acknowledgements in TCP SACK
\cite{Mathis96} or partial acknowledgements in NewReno
\cite{Floyd99} can successfully tackle multiple packet losses
within a single window. Further, more accurate knowledge of the state of the
system (e.g., packets in flight, pending acknowledgements) can help
in the precise control of the injection of packets in the network upon heavy losses,
so that the prescribed congestion window is always kept full and retransmission
timeouts are avoided. Such an extension can be used supportively to the
TCP SACK scheme, as is implemented in the Forward Acknowledgement algorithm \cite{Mathis96b}.
\item[-] Explicit congestion or loss
notification (ECN, ELN, resp. \cite{Balakrishnan98,Ramani00}) can
help to discriminate random losses from those due to congestion,
and thus invoke congestion avoidance and control procedures only
when necessary. 
\item[-] Estimates regarding the amount of backlogged
packets at the receiver queue or estimates of the available
bandwidth (such as the schemes in TCP Vegas, Westwood and Jersey
\cite{Brakmo95,Mascolo01,Xu04}) can assist in configuring the
congestion window more properly for wireless links.
\end{itemize}
Further, we mention a list of good practices, extracted from \cite{Inamura03},
that alleviate many of the problems.
 \begin{itemize}
\item[-] The value of the receive window should be chosen to match the
bandwidth-delay product of a 3G link (about 8-50 KB). 
\item[-] TheTCP timestamps option \cite{Jacobson92} allows more frequent RTT
samples (instead of one RTT per window of data as in normal TCP),
enabling the sender to adapt quicker to changing network
conditions and avoid spurious timeouts. This can be especially
useful for 3G links which experience larger RTT variations,
although an additional overhead is incurred. \item[-] The {\em
Limited Transmit} algorithm \cite{Allman01} can extend TCP's Fast
Retransmit/Fast Recovery by sending a new segment in response to
each of the first two duplicate acknowledgements received. This
keeps data flowing in the system and avoids having to wait for a
3rd {\em dupack}.
\item[-] The MTU discovery procedure \cite{McCann96} can be
implemented, which allows a sender to determine the maximum
end-to-end transmission unit (PDU) supported by the link layer for
a given routing path. This is useful since the default PDU
fragmentation size may be too small, and a larger size can be
tolerated. \item[-] Common applications for Internet-enabled cellular
mobile devices require only the transmission of a few segments'
worth of data. Therefore, it is particularly effective to have a
high initial window, instead of performing slow start. An initial
window of up to four segments is suitable for most applications,
while it has a negligible effect on larger data transfers
\cite{Allman02}.
\end{itemize}
\section{Splitting end-to-end connections}
\begin{figure}[!htp]
    \centering
    \includegraphics[scale=1]{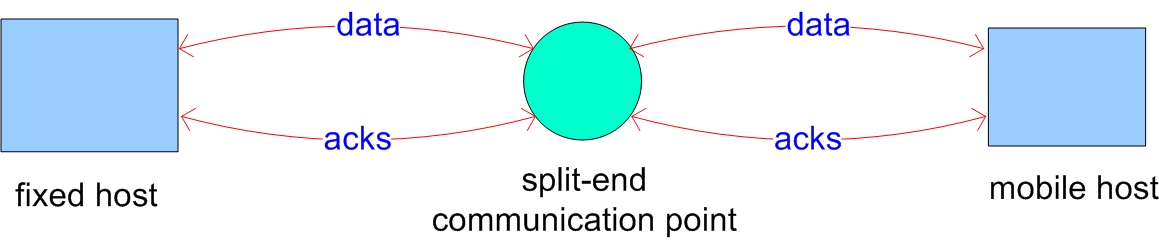}
    \caption{The split-end TCP approach}
    \label{fig3}
\end{figure}

An alternative way of dealing with TCP problems in wireless links
is to ``isolate'' the wireless portion of the path by dividing the
end-to-end connection into wired and wireless connections. This is
the so-called {\em split-end} approach: each connection is
separate, different flow control windows can be maintained, and
acknowledgements are generated for each portion separately.
Further, error control mechanisms, packet sizes and timeouts can
be different for each connection (Fig.~\ref{fig3}).

This approach is reasonable in view of the distinct
characteristics of wired and wireless paths and ensures that the
more problematic wireless TCP behavior has the least impact on the
fixed network. It is also especially convenient for cellular
networks, where the mobile host is only a single hop away from the
fixed network. The point of detachment should be as close to the
mobile host as possible, either at the base station or radio
network controller.

The isolation of the wireless link is the main advantage of a
split-end approach. It also helps in faster reaction to the errors
caused by mobility and wireless transmission, since an end
communication point is maintained very close to the source of
these errors. We can also take advantage of this structure to
simplify the protocol communication between the split end-point
and the mobile host. For example, the TCP header size can be
reduced by cutting down on unused options (e.g., SACK, timestamps,
etc.). Additionally, if IP mobility is not implemented we can
dispense with the overhead of an IP layer altogether.

However, such an approach also has several shortcomings. The need
for intermediate protocol processing incurs extra delay for the
packet transmissions. The approach also prerequisites the
existence of large buffers at the split-end point to absorb
processing and transmission delays. These two issues further raise
the question of scalability, since a base station or controller
may be overwhelmed if it has to serve a large number of mobile
hosts with multiple network connections. Additionally, there can
be a high latency during handoffs and the seamless handover
procedure is obstructed. This issue is further discussed in the
next section which elaborates on terminal mobility.

Finally it must be noted that split-connection approaches violate
end-to-end semantics, as an acknowledgement may be delivered to
the sender before the data actually reaches its destination. This
is disastrous for applications that rely on TCP semantics to
control traffic. However, common applications (e.g., ftp, HTTP,
SMTP) do not face a problem, since they have their own application
layer acknowledgements to control traffic.
\section{More on host mobility}
\subsection{Handovers}
TCP robustness to handovers depends very much on the handover
procedure itself, and issues such as accurate prediction of signal
strength and prompt initiation, duration (in hard handover) and
signal combining (soft handover).

For hard handover\footnote{Inter-frequency handover and
inter-system handover are also hard handovers.}, even though care
is taken from the part of the radio network controller to complete
the procedure as soon as possible, a surplus of TCP problems will
inevitably occur. Reasons are transmission losses, the induced
latency, but also buffer overflow, since the sending and receiving
of data is suspended until the handover is completed, whilst a
fixed host continues to send data as the whole procedure is
normally transparent to it.

In a CDMA network where soft handover\footnote{This is always an
intra-frequency handover. More information on handovers in 3G
systems can be acquired through a very educational thesis
\cite{Chen03}.} is more easily implemented (although complexity is
still an issue), errors can be kept to a minimum. However,
problems may still occur if handover is not initiated promptly; it
should be noted that signal fluctuations are difficult to predict
and it is hard to obtain optimized handover margins. In another
extent, the implementation of soft handover may not be purely
advantageous, since the transmission and reception of multiple
signals can burden the fixed links, especially if a mobile is
connected to more than two base stations. Moreover, transmission
errors may occur due to an imperfect power control mechanism
during soft handover, as it is much more difficult to reach a
perfect scheme in this case, while achieving the desired
macrodiversity gain. Last but not least, the decision phase in CDMA soft handover
suffers from additional delay, due to the need for timing synchronization
of the signals from different base stations.

In split-end approaches further problems are encountered if a
connection is split at the base station. Normally, all information
pertaining to a TCP connection (send and receive windows, sequence
numbers, retransmission status and timers, etc.) established on
behalf of a mobile host must be handed over to the new BS. This
shift of information burdens the network and increases the
latency. This problem stipulates that it is better to split the
connection at the radio network controller, as changes to
different RNCs will occur much more rarely. The requirement for
shift of information also brings about the inability to implement
soft handover when the connection is split at the base station.

Apart from designing the procedure itself as best as possible, an
extra step that can be taken to improve TCP handover performance
is to freeze the retransmission timers during the handover
procedure, or suppress (or suspend) the sending of duplicate
acknowledgements until it is completed.

Finally, a method to eliminate handover latency in the downlink is
multicasting. The basic principle is as follows: The system is
informed to anticipate handoffs and multicasts data destined to
the mobile host to nearby base stations in advance. Hence the goal
is essentially the same as with the soft handover scheme, but the
implementation is different (formation of multicast groups, etc.)
and it usually operates on a larger scale (more BS are involved,
which makes it more suitable for picocellular networks). There
exist many schemes that propose multicasting, interested readers
can for instance look up \cite{Ghai94,Seshan96}.
\subsection{IP mobility}
While the radio access handover and link layer mobility are provided by the CDMA
access scheme, the 3G core network needs to be able to support IP handover
for unabridged connectivity to the mobile host.

Cornerstones for IP mobility support are laid with the {\em Mobile IP} scheme \cite{Perkins96}.
Mobile IP solves the problem of mobility of a node by managing the correspondence between
the changing IP address of a mobile host, called the {\em care-of address}, and the {\em home address}
permanently or semi-permanently assigned to the host. All packets sent to a mobile node's home
address are tunnelled by a home agent to the current care-of address of the mobile.

Mobile IP demands the integration of mobility management entities into the core 3G architecture.
However, the general problem concerning TCP/IP transfers is to resume the sending and reception of IP packets as soon as possible 
subsequent to a handover, with the least disruption to the transmission sequence \cite{Koodli01}.
During handoff additional latency which deteriorates TCP behavior is incurred by procedures for:
1)\: address resolution delay and
2)\: home network registration.

There exist proposed schemes that take action against  the above deficiencies.  {\em Hierarchical mobile IP}
\cite{Soliman05} aims to reduce the home network registration delay by implementing a hierarchical 
mobility management scheme, where short-range mobility is handled through a local router called 
{\em mobility anchor point} (MAP). When a mobile moves within a MAP domain, handover latency is greatly 
reduced as the binding update needs to reach the MAP only. Secondly, fast (or low-latency) handover \cite{Koodli05}
aims to reduce the address resolution delay by address pre-configuration. It involves cooperation between the old 
and the new access router so that the new address can be resolved and packets can start being forwarded to 
the new access router before the handover is executed. Finally, there also exist multicast approaches for IP mobility (e.g., \cite{Helmy00}). 

A critical note on the fast handover approach is that packet forwarding results in a lot of
re-ordering at the receiver, since packets through the new access router may arrive sooner at their destination.
In that case, selective TCP acknowledgement schemes are called for to assist recovery (the relatively slow and 
loss-neglecting recovery procedure in Reno and NewReno turns out to be insufficient and these schemes can 
perform worse than Tahoe, see  \cite{Jaiswal04}).

Clearly, one can combine the key benefits of these approaches into a new mechanism. A scheme that builds on
top of the hierarchical and the fast handover approach --and also employs a multicasting function) is S-MIP 
(Seamless Mobile-IP \cite{Hsieh03}). This achieves fewer packet losses and more accurate handover through 
movement prediction, improving  the overall behavior. S-MIP also keeps re-ordering problems to a minimum by
maintaining different buffers for forwarded packets, which are flushed out before packets from the new access router.
The integration of S-MIP mechanisms into the UMTS architecture has been proposed in \cite{Lau05}, exhibiting 
superior performance.
\section{A summary of major points}
This review has offered an analytic presentation of performance problems and solutions related to the transfer of TCP data 
flows in networks including wireless paths, with emphasis on 3G cellular links. Certain major points can be highlighted.

First, TCP performance problems can be kept to a minimum by appropriate changes in the end-to-end
protocol mechanisms. Especially the use of existing experimental schemes, such as selective or partial acknowledgements and
explicit loss or congestion notifications greatly improve performance. But also the more accurate tracking of the state 
and available bandwidth of a link are important ingredients of an improved scheme.

Moreover, the addition of link level mechanisms in the ``last mile'' can help in faster reaction and recovery from losses and timeouts. 
This is requisited and practically more helpful, since it is very difficult for end-host protocol modifications to be uniformly supported in a network.
Maximization of performance is then achieved by using a TCP-aware link layer protocol. 
In this case however, it is important to understand how the competitive
ARQ mechanisms in the TCP and link layer interact with each other to affect overall performance and how the specific 
parameters of each should be fine-tuned.

On what concerns the split-end approach, dividing the end-to-end connection into wired and wireless paths is reasonable 
and seems especially attractive in single-hop networks, but it is not strictly necessary. As was shown in \cite{Balakrishnan97}, 
although an appropriate split-end scheme can perform better than the standard TCP Reno, its performance can be worse than that of an advanced end-to-end or a well-tuned TCP-aware link layer protocol.
 
It should also be stressed that the complete exploitation of Internet services over 3G links is not accomplished unless IP mobility is supported. 
Although this demands the integration of IP mobility management entities into the core 3G architecture and incurs 
additional TCP problems, in the long run an all-IP cellular network will provide for cost savings and ease of use.

Finally, coupled with TCP performance are issues concerning appropriately large buffer sizes to combat link assymetry and achieve good utilization in 3G cells, and the design of fast and rate-preserving scheduling algorithms to alleviate problems occuring by the intermittent allocation of resources in such systems.

\begin{thebibliography}{37}
\bibitem{Korhonen03} J.~Korhonen, {\em Introduction to 3G Mobile
Communications}, 2nd ed., Artech House Publishers, 2003.
\bibitem{Stevens94} R.~Stevens, {\em TCP/IP Illustrated (Vol.~1)}, Addison--Wesley, Reading, MA, 1994.
\bibitem{Jacobson88} V.~Jacobson, ``Congestion Avoidance and
Control'', {\em Proc. ACM Sigcomm 1988}, Stanford, CA.
\bibitem{Postel81} J.~Postel, ``Transmission Control Protocol --- DARPA Internet Program Protocol
Specification'', RFC 793, September 1981.
\bibitem{Stevens97} W.~Stevens, ``TCP Slow Start, Congestion
Avoidance, Fast Retransmit and Fast Recovery Algorithms'', RFC
2001, January 1997.
\bibitem{Inamura03} H.~Inamura, G.~Montenegro, R.~Ludwig,
A.~Gurtov, F.~Khavizov, ``TCP over Second (2.5G) and Third (3G)
Generation Wireless Networks'', RFC 3481, February 2003.
\bibitem{Tian05} Y.~Tian, K.~Xu, N.~Ansari, ``TCP in Wireless
Environments: Problems and Solutions'', {\em IEEE Radio
Communications}, March 2005, pp. 27--32.
\bibitem{Thoppian} M.~Thoppian, A.~Veduru, ``TCP for Wireless
Networks'', http://utdallas.edu/$\sim$sudha/TCP\_Wireless.pdf
\bibitem{Parkvall01} S. Parkvall, E. Dahlman, P. Frenger, P. Beming, M. Persson, ``The high speed packet data evolution of WCDMA'',
{\it{Proc. 12th IEEE PIMRC}}, San Diego, USA, 2001.
\bibitem{Bender00} P. Bender, P. Black, M. Grob, R. Padovani, N. Sindhushayana, A. Viterbi,
``CDMA/HDR: A bandwidth-efficient high-speed wireless data service
for nomadic users'', {\it IEEE Communications Magazine}, pp. 70--77,
July 2000.
\bibitem{Jacobson90} V.~Jacobson, ``Compressing TCP/IP Headers for Low-Speed Serial Links'',
RFC 1144, February 1990.
\bibitem{3GPP2} 3GPP2, ``Data Service Options for Spread Spectrum
Systems: Radio Link Protocol Type 3'', Specification C.S0017-010-A
Ver. 2.0, September 2005.
\bibitem{3GPP} 3GPP, ``Radio Link Control (RLC) protocol
specification'', Ver. 6.5.0, 2005.
\bibitem{Mathis96} M.~Mathis, J.~Mahdavi, S.~Floyd, A.~Romanow,
``TCP Selective Acknowledgement Options'', RFC 2018, October 1996.
\bibitem{Floyd99} S.~Floyd, T.~Henderson, ``The NewReno
Modification to TCP's Fast Recovery Algorithm'', RFC 2582, April
1999.
\bibitem{Mathis96b} M.~Mathis, J.~Mahdavi, ``Forward Acknowledgment: Refining TCP Congestion Control'',
{\em Proc. ACM Sigcomm}, Palo Alto, California, USA, August 1996, pp. 281--291.
\bibitem{Ramani00} R.~Ramani, A.~Karandikar, ``ECN in TCP over Wireless Networks'',
{\em IEEE Int'l Conf. on Personal Wireless Communications''},
2000.
\bibitem{Balakrishnan98} H.~Balakrishnan, R.~Katz, ``Explicit Loss
Notification and Wireless Web Performance'', {\em IEEE Globecom
Internet Mini-Conference}, Sydney, Australia, November 1998.
\bibitem{Brakmo95} L.~Brakmo, L.~Peterson, ``TCP Vegas: End to End
Congestion Avoidance on a Global Internet'', {\em IEEE J. Selected
Areas in Commun.}, vol.~13, no.~8, October 1995, pp. 1465--1480.
\bibitem{Mascolo01} S.~Mascolo, C.~Casetti, M.~Gerla,
M.Y.~Sanadidi, R.~Wang, ``TCP Westwood: Bandwidth Estimation for
Enhanced Transport over Wireless Links'', {\em ACM Mobicom '01},
pp. 287--297, July 2001.
\bibitem{Xu04} K.~Xu, Y.~Tian, N.~Ansari, ``TCP-Jersey for Wireless
IP Communications'', {\em IEEE J. Selected Areas in Commun.},
vol.~22, no.~4, May 2004, pp. 747--756.
\bibitem{Jacobson92} V.~Jacobson, R.~Braden, D.~Borman, ``TCP
Extensions for High Performance'', RFC 1323, May 1992.
\bibitem{Allman01} M.~Allman, H.~Balakrishnan, S.~Floyd,
``Enhancing TCP's Loss Recovery Using Limited Transmit'', RFC
3042, January 2001.
\bibitem{McCann96} J.~McCann, S.~Deering, J.~Mogul, ``Path MTU
Discovery for IP version 6'', RFC 1981, August 1996.
\bibitem{Allman02} M.~Allman, S.~Floyd, C.~Partridge, ``Increasing
TCP's Initial Window'', RFC 3390, October 2002.
\bibitem{Chen03} Y.~Chen, {\em Soft Handover Issues in Radio
Resource Management for 3G WCDMA Networks}, PhD Thesis, Queen
Mary, University of London, September 2003.
\bibitem{Ghai94} R.~Ghai, S.~Singh, ``An architecture and communications protocol for picocellular networks"
{\em IEEE Personal Communications}, 1994.
\bibitem{Seshan96} S.~Seshan, H.~Balakrishnan, R.~Katz, "Handoffs in cellular wireless networks: The Daedalus
implementation and experience" {\em Wireless Personal Communications}, vol.~4, no.~2, March 1997, pp. 141--162.
\bibitem{Perkins96} C.E.~Perkins, ``IP mobility support'', RFC 2002, October 1996.
\bibitem{Koodli01} R.~Koodli, C.E.~Perkins, ``Fast Handovers and Context Transfers in Mobile Networks'',
{\em ACM Sigcomm Computer Communication Review}, vol. 31, no. 5, pp. 37--47, October 2001.
\bibitem{Soliman05} H.~Soliman, C.~Castelluccia, K.~El Malki, L.~Bellier, ``Hierarchical Mobile IPv6 Mobility Management'',
RFC 4140, August 2005.
\bibitem{Koodli05} R.~Koodli, ``Fast Handovers for Mobile IPv6'', RFC 4068, July 2005.
\bibitem{Helmy00} A.~Helmy, ``A Multicast-Based Protocol for IP Mobility Support'', {\em Proc. NGC 2000 on Networked 
Computer Communication}, Palo Alto, USA, November 2000, pp. 49--58. 
\bibitem{Jaiswal04} S.~Jaiswal, S.~Nandi, ``Simulation-Based Performance Comparison of TCP-variants over Mobile 
IPv6 Based Mobility Management Schemes'', {\em 29th IEEE Conference on Local Computer Networks} (LCN 2004), Tampa, Florida, USA, November 2004.
\bibitem{Hsieh03} R.~Hsieh, Z.-G.~Zhou, A.~Seneviratne, ``S-MIP: A Seamless Handoff Architecture for Mobile IP'',
{\em Proc. IEEE Infocom 2003}, San Fransisco, USA, March 2003, pp. 1774--1784.
\bibitem{Lau05} C.-K.~Lau, ``Improving Mobile IP Handover Latency on End-to-End TCP in UMTS/WCDMA Networks'',
{\em Proc. CoNEXT'05}, Toulouse, France, October 2005.
\bibitem{Balakrishnan97} H.~Balakrishnan, V.N.~Padmanabhan, S.~Seshan, R.H.~Katz, 
``A Comparison of Mechanisms for Improving TCP Performance over Wireless Links'', {\em IEEE/ACM Transactions on Networking},
vol. 5, no. 6, pp. 756--769, December 1997.
\end{thebibliography}
\end{document}